# A FRAMEWORK FOR A SMART SOCIALBLOOD DONATION SYSTEM BASEDON MOBILE CLOUD COMPUTING

Almetwally M. Mostafa

College of Computer and Information Sciences, King Saud University, Riyadh, KSA
&Faculty of Engineering, Alazhar University, Cairo, Egypt

Ahmed E. Youssef

College of Computer and Information Sciences, King Saud University, Riyadh, KSA
& Faculty of Engineering, Helwan University, Cairo, Egypt

GamalAlshorbagy

College of Computer and Information Sciences, King Saud University, Riyadh, KSA

## ABSTRACT

*Blood Donation and Blood Transfusion Services (BTS) are crucial for saving people's lives. Recently, worldwide efforts have been undertaken to utilize social media and smartphone applications to make the blood donation process more convenient, offer additional services, and create communities around blood donation centers. Blood banks suffer frequent shortage of blood;hence, advertisements are frequently seen on social networks urging healthy individuals to donate blood for patients who urgently require blood transfusion. The blood donation processusuallyconsumesa lot of time and effort from both donors and medical staff since there is no concrete information system that allows donorsand blood donation centers communicate efficiently and coordinate with each other tominimize time and effort required for blood donation process. Moreover, most blood banks work in isolation and are not integrated with other blood donation centers and health organizations which affect the blood donation and blood transfusion services' quality. This work aims at developing a Blood Donation System (BDS) based on the cutting-edge information technologies of cloud computing and mobile computing. The proposedsystem facilitates communication between blood donorsand blood donation centers and integrates the blood information dispersed among different blood donation centers and health organizations acrossa country.Stakeholders will be able to use the BDS as an application installed on their smartphones to help them complete the blood donation process with minimal effort and time. Thisapplication helps people receive notifications on urgent blood donation calls, know their eligibility to give blood, search for the nearest blood center, and reserve a convenient appointment using temporal and/or spatial information. It also helps establish a blood donation community through social networks such as Facebook and Twitter.*

                                                                                            1

## 1. INTRODUCTION

The ever-increasing number of trauma patients, particularly those involved in car accidents, major surgeries, chemotherapy for malignant diseases, and patients for long-term blood therapy, such as those with sickle cell anemia and thalassemia has increased the demand for blood in many countries [1]. Over the last three decades, the source of blood has shifted dramatically from imported blood to local blood donation. Blood banks found in different hospitals take care of the entire donation process, including recruiting donors, collecting and screening of the donated blood and preparation and storage of blood components. At the present time, the source of donated blood is a combination of involuntary donors (relatives, friends, and workmates), and a voluntary non-remunerated donors usually through campaigns. Shortage in voluntary, non-remunerated blood donors poses major challenges to Blood Transfusion Services (BTS); therefore, continuous efforts are made to attract healthy members of the public to become voluntary, non-remunerated blood donors. Studies have been identified numerous factors that would influence the recruitment and the retention of blood donors including a wide range of psychological, physiological and sociodemographic factors [2]. A recent psychology study [1] on the attitude to blood donation among male students at King Saud University (KSU) reflected an encouraging strong positive attitude toward blood donation. The study highlighted the need to invest in awareness and motivationson blood donation through campaigns so that current donors will continue donating and non-donors will be encouraged to begin donating. Currently, blood campaigns rely on frequent advertisements on different types of media such as TV, newspapers, radio stations, and communication through social networks such as Facebook and Twitter. For example, figure 1 show an advertisement (in Arabic) posted on SaudiNews50 twitter account that invites people to donate blood for an emergency case existed at King Fahd Medical City. This request has been rewetted 158 accounts to their followers which mean that thousands of people will be aware with this case.

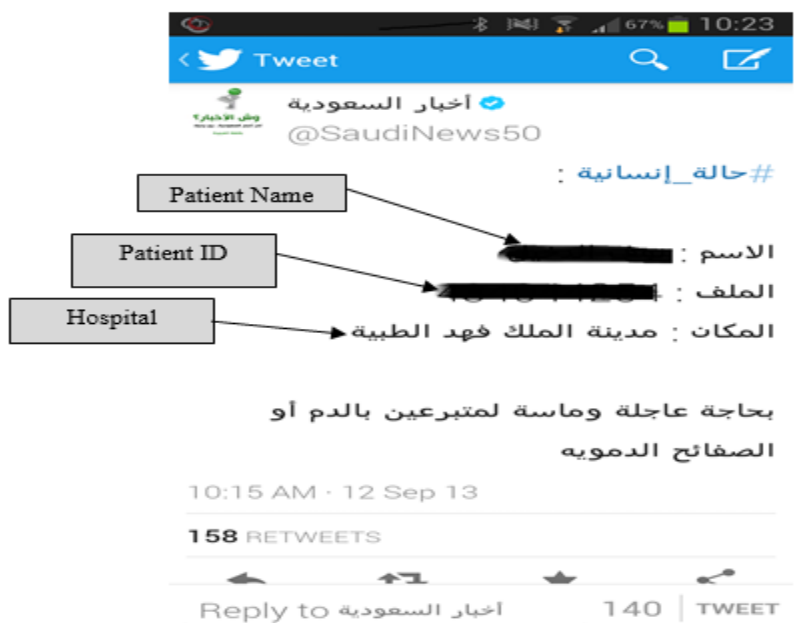

Fig. 1: Screen shot for emergency blood donation request on Twitter

Although these advertisements would help increase voluntary blood donation, we see several problems that would obstacle the goal of these advertisements such as:

1. Awareness messages are blindly sent to too many people regardless of their current location and eligibility to donate in terms of proper blood type, health status, time between successive donations, etc. We need to send this message to a person who is nearby and eligible to donate.
2. No communication between donors and blood bank(s) that accept their donation. We need an efficient and effective way that helps donors and blood banks communicate and coordinate with each other to minimize donation effort and time due to large number of donors arriving at the donation center at the same time.
In addition to lack of effective communication channels between donors and blood donation center, blood donation centers work in isolation with no mechanism for exchanging blood information. Currently, this information is scattered among different databases in blood donation centers, health organizations, and social networks such as Facebook and Twitter. What is needed is an innovative information system that integrates this information and facilitates communication among different stakeholders (donors, donation centers, and health organizations).

This work aims at developing a Blood Donation System (BDS) based on the cutting-edge information technologies of cloud computing and mobile computing. The proposed system facilitates communication between blood donors and blood donation centers and integrates the blood information dispersed among different blood donation centers and health organizations across the country. Stakeholders will be able to use the BDS as an application installed on their smartphones to help them complete the blood donation process with minimal effort and time.

This application helps people receive notifications on urgent blood donation calls, know their eligibility to give blood, search for the nearest blood center, and reserve a convenient appointment using temporal and/or spatial information. It also helps establish a blood donation community through social networks such as Facebook and Twitter.An up to date donor profile with information on his/hercurrent location,his blood type, last donation date, etc, will be used recorded on his/her page. Information on blood donation needs will be smartly passed to appropriate donors which helps find a nearby appropriate donor at the appropriate time. The rest of this paper is organized as follows in section 2 we present related work,in section 3 we discuss the concept of pervasive cloud computing, in section 4 we present our research methodology, in section 5 we introduce our proposed Blood Donation System, and in section give our conclusion and future work.

## 2. RELATED WORK

The World Health Organization (WHO) has announced that mobile-health (m-health) has the "potential to transform the face of health service delivery across the globe" [3]. The evidence-base on feasibility application and user preferences for mobile health application is relatively limited and nascent. Mobile phone-based BTS studies began in the past few years and mobile-BTS is even more innovation. The potential impact of mobile or self-management application is dependent on their scalability and adaptability to be acceptable and attractive to diverse user groups and priorities.Masser et al. [13] provides an overview of existing research in this area. Brionesa et al. [14] who studied social media usage of the Red Cross in the United States concluded that services such as Facebook and Twitter are used by the organization to support and develop relationships focused on recruiting and maintaining volunteers, updating the community on disaster preparedness and response, and engaging the media. While the American Red Cross maintains a dedicated Facebook page to share stories about donors and blood donations to engage the community, LifeSouth Community Blood Centers in Florida created a Facebook application "I give Blood" which allows donors to share their donation history with friends. SocialBlood[18] provides a web service utilizing web and Facebook to create connections between blood donors and blood transfusion receivers anduses location-aware technology to find donors nearby in the event of an emergency. Sahlgrenska University Hospital in Gothenburg, Sweden has launched a massive social media campaign to encourage blood donors to help meet a shortage [4].

Besides utilizing social media services, various smart phone applications have been released to the public around the topic of blood donations. In India, a smart phone based virtual blood bank has been proposed [5]. This system uses General Packet Radio Service (GPRS) and a centralized server to store blood donors and blood banks information. People who seek blood also communicate with the server through their mobile devices, specifying their blood type and current location in a subscriber application. The server matches the blood type and location with the profiles of registered donors or blood banks, retrieves the information and sends it to the seeker via GPRS. Moreover, the Indian Medical Association (IMA) has launched a Short Message Service (SMS) for blood bank services [6]. In Bangladesh, a quickly access blood donation service to assist in the management of blood donor records has been proposed [7]. Asian Development Bank cooperation with Microsoft during annual meeting held in Delhi, India, has announced the winning of mobile apps that aims to manage blood donor networks [8]. The Swiss and the German Red Cross have been adopted web portal and social network to help users to search and locate the proper blood donation centers. Blood Alliance, which serves hospitals in North-East Florida and portions of Georgia and South Carolina has launched iDon8, an application that offers basic scheduling options as well as a visualization of available blood stock [9]. In Ireland [10] a smart phone application enabling existing and potential donors to monitor the national blood levels by blood type.

Dutch Institute for Healthcare Improvement [11] suggests the use of information technology for enhancing of blood transfusion to patients. The report recommends recording donor data, patient data, and other related datato make it possible to monitor transfusion history record of patient as well as donor. Electronic recording of such data reduces nurses and technicians for performing transfusion preparations. Marcus et al. [12] explored that some donors would like to trace their blood donation. One interviewed said: "The Red Cross has a deep knowledge about what they are doing. If they share this, I would be more likely to engage. I like understanding systems." Other interviewed said: "Would like to know more about the geographical journey my blood took." Electronic registration for blood donation systems is crucial for improving safety of blood transfusion system. Blood donation and transfusion registration system, facilitates monitoring and displaying the local blood stock levels for the donor/patient's area. Registration also helps visualize donations on a city map as well as visualization of the pervasiveness of different blood groups across a city. The focus of this work lies on the automation and integration of blood donation systems, blood bank systems, health organization systems, and social media for the sake of motivating people to donate blood and facilitate blood donation process. To the best of our knowledge, such apps and systems have not been subjected to any academic scrutiny to date.

## 3. MOBILE CLOUD COMPUTING

In recent years, mobile devices have started becoming abundant with applications in various categories such as entertainment, health, games, business, social networking, travel and news. The reason for the increasing usage of mobile computing is its ability to provide a tool to the user when and where it is needed regardless of user movement, hence supporting location independence. Indeed, 'mobility' is one of the characteristics of a mobile computing environment where the user is able to continue his/her work seamlessly regardless of his/her movement [15]. However, exploiting its full potential is difficult due to some inherent problems such as limited scalability of users and devices, limited availability of software applications and information, resources scarceness in embedded gadgets and frequent disconnection and finite energy of mobile devices. As a result, a wide range of applications are difficult to run in mobile devices. These applications fall into different areas such as image processing and recognition, text translation, speech recognition, multimedia, social networking, and sensor data applications

Cloud Computing (CC) platforms possess the ability to overcome these discrepancies with their scalable, highly available and resource pooling computing resources. CC platform is an aggregation of hardware, systems software and application software where the application software is delivered as a service over the Internet, Software-as-a-Service (SaaS), and the systems software and the hardware in data centers provide Platforms-as-a-Service (PaaS), and Infrastructure-as-a-Service (IaaS). The main idea behind CC is to offload data and computation to a remote resource provider (the Cloud or the Internet) which offers many interesting characteristics such as: on-demand self-service, broad network access, resource pooling, rapid elasticity, multitenancy, and scalability [16].

The concept of offloading data and computation in the Cloud, is used to address the inherent problems in mobile computing by using resource providers (i.e., cloud resources) other than the embedded devices themselves to host the execution of user applications and store users' data. The problems are addressed as follows: 1) by exploiting the computing and storage capabilities (resource pooling) of the cloud, mobile intensive applications can be executed on low resource and limited energy mobile devices, 2) the broad network access of the cloud overcomes the limited availability and frequent disconnection problems since cloud resources are available anywhere and at any time, 3) the infrastructure of cloud computing is very scalable, cloud providers can add new nodes and servers to cloud with minor modifications to cloud infrastructure, therefore; more services can be added to the cloud, this allows more mobile users to be served and more portable devices to be connected.

The study by Juniper Research, which states that the consumer and enterprise market for cloud-based mobile applications is expected to rise to $9.5 billion by 2014 [17]. In the paper, we investigate how interesting characteristics of mobile cloud computing can be deployed to develop a new generation of Blood Donation Systems that reduce time and effort consumed in blood donation process by providing efficient communication channels between donors and blood donation centers and integrates blood information dispersed among different blood centers, health organizations, and social media.

## 4. THE PROPOSED FRAMEWORK

The Blood Donation System (BDS) is an information system that deploys mobile computing empowered by cloud computing to provide efficient blood donation services. BDS is comprised of two main components as shown in Fig.2, the Cloud Computing (CC) component and the Mobile Computing (MC) component.

The CC Component: this component represents the cloud with all its service models (IaaS, PaaS, SaaS). It provides stakeholders (donors, blood bank staff, health organizations staff) in the mobile computing environment with all blood donation services. These services include:

1. Ontology interface system
2. Emergency Service provided by national/ regional donors database
3. Blood donation campaign service
4. Blood Donation Registration Service
5. Blood Donation Reservation Service

The MC Component: this component represents the mobile environments where blood donation stakeholders interact and request services from the cloud component. It provides interfaces between the users and the cloud in the form of smartphone blood donation application. This application enables users to receive blood donation alerts in real time fromblood banks and health organizations. In addition, critical and emergency calls can be displayed by illegible users who are in urgent need for blood transfusion. The MC

Component includes the following entities:

1. Service Directory (SD): which is a directory for all blood donation services

2. User Agent (UA): this refers to any smartphone or mobile device that receives service equest from the user and sends it to SD. It is an interface between the mobile user and SD. When the SD receiver a service request from a user, it forwards this request to the appropriateservice unit in the cloud.

The proposed framework aims at bring the following benefits in terms of blood donation services:

1. Provide fast and smart response actions in emergency cases that help find the appropriate donor immediately
2. Monitor blood repository by its sources and locations and provide information for health organizations on blood availability all over the country
3. Increase blood donation loyalty in the community by saving donors and medical staff time and effort consumed in blood donation process

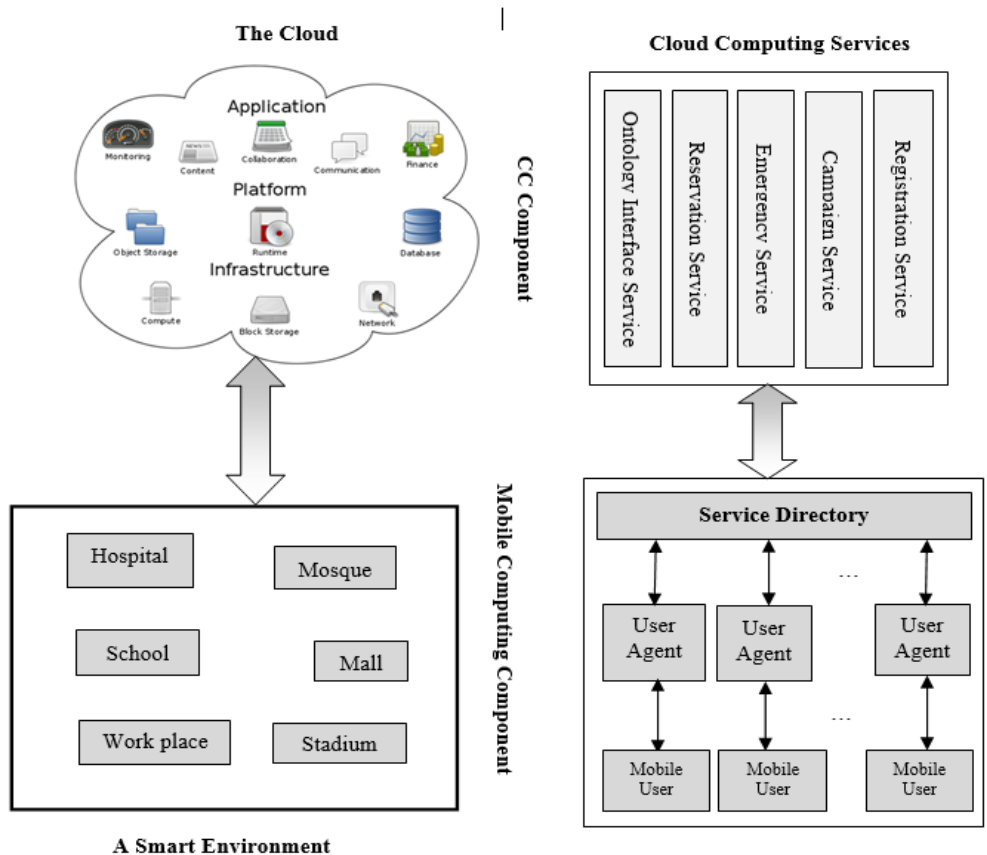

Fig.2: The Frameworkfor BDS

## 5. BLOOD DONATION SYSTEM ARCHITECTURE

Figure 3 shows the environment of BDS, the system brings together different type of stakeholders including blood donors, blood bank staff, and health organization staff through the mobileApp. On the other hand, BDS cooperates with other standalone external systems such as blood bank and hospital databases. In addition, the proposed BDS is able to connect to social networks such as Facebook and Twitter as well as national and international health organizations. The system is comprised of the following interconnected functional components:

1. Ontology interface: provides a smart search for suitable donor(s)
2. Emergency Blood Donation: provides a fast assistance for people who are in urgent need for blood.
3. Campaign Organizer:helps individuals, groups of users, blood donation centers, health organizations, or ministry of health effectively organize and manage a blood donation campaign.
4. Registration: provides registration services for all stakeholders of the system (blood banks, donors, health organization)
5. Reservation: allows donor reserve appointments for donation. This will relieve both blood bank staff and donors from wasting their time and exerting extra effort in blood donation process.

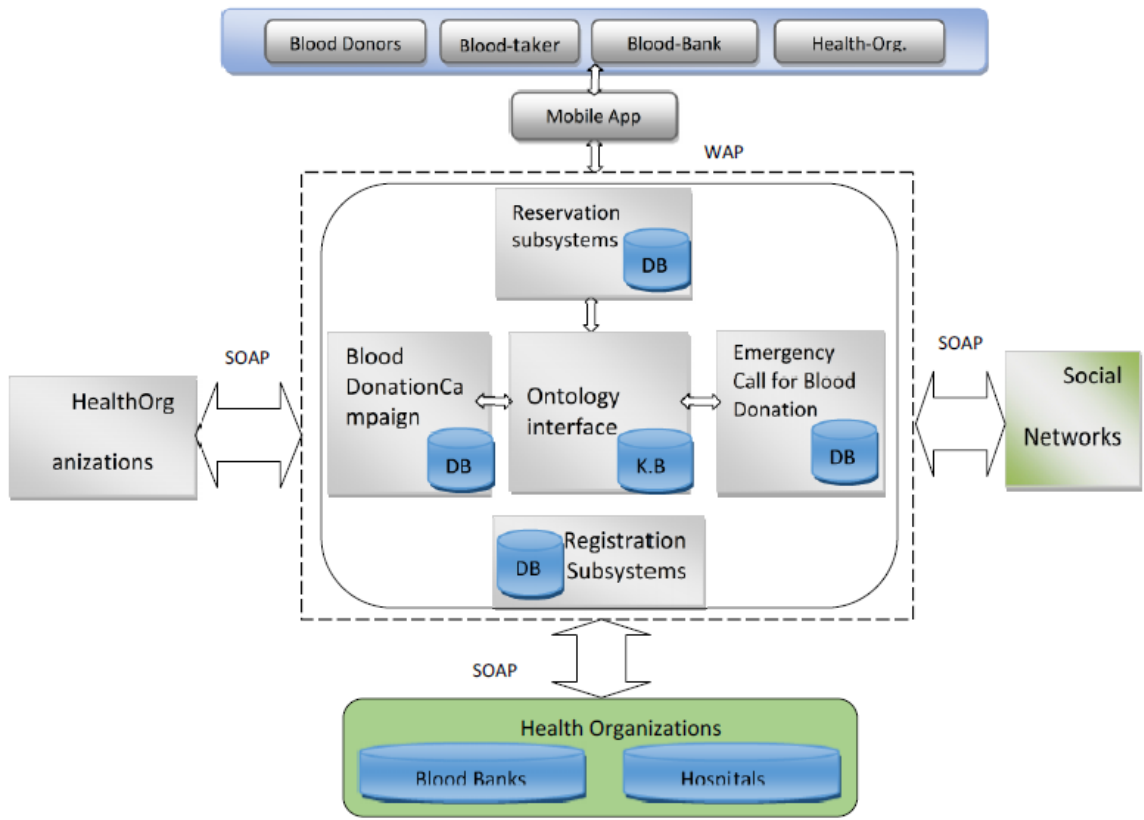

Fig.3: BDS Architecture

The interaction between different entities helps acquire information on blood bank and donors' current situations, predict the future needs, and take proactive actions. Integration with health organizations helps obtain statistical information such as: how many blood campaigns they initiated in the country, what regions covered by these campaigns, and how many donors participated. Integration with blood donation centers allows medicalstaff sends invitation for blood donation on behalf of patient directly and invite appropriate donors to be ready in the right place and right time. Appropriate donors are those who have proper blood type and are eligible to donate. Integration with social networks provides important information on patients and donors such as current location, their relatives, friends, and followers. Direct social relationships between individuals are improved dramatically. Every donor knows who takes his/her blood and every patient knows who gave him/her blood.

## 6. CONCLUSION AND FUTURE WORK

In paper we proposed a Blood Donation System (BDS) based on the cutting-edge information technologies of cloud computing and mobile computing. The proposed system facilitates communication between blood donors and blood donation centers so that the appropriate donor can be reached just on time. It also integrates the blood information scattered among different blood donation centers and health organizations across the country to improve blood donation service quality. Stakeholders are able to use the BDS as an application installed on their smartphones to help them complete the blood donation process with minimal effort and time. This application helps people receive notifications on urgent blood donation calls, know their eligibility to give blood, search for the nearest blood center, and reserve a convenient appointment. It also helps establish a blood donation community through social networks such as Facebook and Twitter. In the future, we plan to develop and implement the proposed system in Kingdom of Saudi Arabia (KSA) and examine its effect on blood donation process and blood transfusion service in blood donation centers and health organizations across the Kingdom.